# Large deflections and stability of spring-hinged cantilever beam

Milan Batista

University of Ljubljana, Slovenia

milan.batista@fpp.uni-lj.si

(Nov-Dec 2018)

**Abstract**

In the article, we investigate the influence of spring on the large deflections and stability of a spring-hinged cantilever subject to conservative tip force. Using the closed form solution of the equilibrium equation and closed form solution of Jacobi accessory equation, we determine the beam equilibrium forms and their stability. Also, the solution for spring-hinged cantilever bema subject to a follower force is given. Results are present in the graphical and the tabular form.

*Keywords.* Elastic beams; elastic support; large deformations; stability; Jacobi test;

## 1 Introduction

Cantilever beam represents one of the most common construction element in mechanical and civil engineering [1-3], and in recent decades also in robotics [4], and in micro- and nano-engineering [5-8]. Therefore, because of its importance, the study of the large deflection and stability of a cantilever beam has attracted numerous researches. Most of the works are devoted to clamped cantilever beam (see [9-14] and reference therein). For the spring-hinged cantilever beam subject to a conservative force the literature is not extensive. A discussion of the stability of a spring-hinged cantilever beam is given in books [15, 16] where one can find the derivation of the formula for the buckling force using Euler's method. Rao and Raju [17] analyses the post-buckling behavior of the spring-hinged cantilever beam using the finite element method. Ohtsuki and Yasui [18] solve the large deflection of the spring-hinged cantilever beam under inclined force using elliptic integrals. These authors enhance their calculations with bending tests. Rao and Raju [19] calculate critical load parameter for the cantilever under axial force and distributed load using semi-analytic Rayleigh-Ritz method. Another possible force acting on the cantilever is a follower force on which, especially

24/12/2018 08:50

for pure compressive one, there are different opinions [20-22]. Large deflections of a spring-supported cantilever subject to follower force using elliptic integrals were considered by Rao et all [23]. Rao and Rao [24] examine large deflections of a spring-hinged tapered cantilever beam subject to a rotational distributed loading using Runge-Kutta numerical integration. Shvartsman [25] considers large bending of a spring-supported cantilever subject to follower force using numerical integration. For treatment of stability of the cantilever beam under a follower force using dynamical methods, we refer to [26, 27] and especially for spring-hinged cantilever beam to [28-31]. For other elastically supported cases of beams, see [32-34].

From the available literature, we conclude that the stability of post-buckling forms of the spring-hinged column beam has not yet been reported. Therefore, in this study, we aim to fill this gap. In the next section, we set up the governing equations of the problem using the principle of minimum total potential energy. The first variation of this energy leads to Euler equilibrium equation and its second variation to Jacobi accessory equation [35] which is used for accessing the equilibrium stability. In Section 3 we give the solution of these equations in terms of Jacobi elliptic function. Then in Section 4 we provide some numerical examples and discuss the cantilever in compression in some details. Cantilever subject to a follower force is discussed in Section 5. Finally, the results are summarized in the last section.

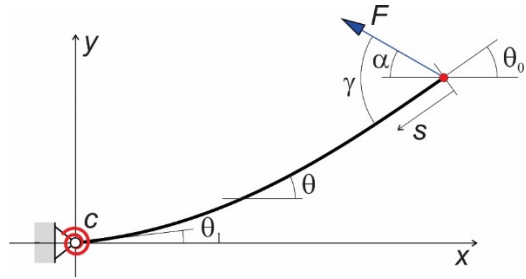

**Figure 1.** Geometry and load of the spring-hinged cantilever beam

## 2 Problem formulation

We consider an elastic spring-hinged cantilever beam subject to force $F$.. The cantilever length is $\ell$, its flexural rigidity is $EI$, the rotational spring stiffness is $c$, and the force inclination angle is $\alpha$ (Fig 1).

*2.1 Basic equations*

The differential equations of the column base curve are

$$\frac{dx}{ds} = -\cos\theta, \quad \frac{dy}{ds} = -\sin\theta \tag{1}$$

in which $0 \leq s \leq \ell$ is the arc length measured from column free end to its fixed end, $x$, $y$ are the base curve coordinates, $\theta$ is the tangent angle. The conditions at the fixed end are $x(\ell) = y(\ell) = 0$. Using this we obtain from (1) the coordinates $x_0 \equiv x(0)$ and $y_0 \equiv y(0)$ of the free end

$$x_0 = \int_0^\ell \cos\theta\, ds, \quad y_0 = \int_0^\ell \sin\theta\, ds. \tag{2}$$

The expression for the cantilever total potential energy $\Pi$ is

$$\Pi = \int_0^1 \tfrac{1}{2} EI\kappa^2 ds - F\cos\alpha\left(\ell - x_0\right) - F\sin\alpha\, y_0 + \tfrac{1}{2} c\, \theta_1^2 \tag{3}$$

where $\theta_1 \equiv \theta(\ell)$ and $\kappa$ is the base curve curvature given by

$$\kappa = -\frac{d\theta}{ds} \tag{4}$$

For the equilibrium, $\Pi$ is to be minimum [16]. This means that the first variation of $\Pi$ must vanish and the second variation of $\Pi$ need to be positive. Below, we will for the derivation of the governing equations of the problems follows the well-known variational procedure [35].

The first variation of $\Pi$ as given by (3) is, using (2),

$$\delta\Pi = \int_0^\ell \left[EI\kappa\,\delta\kappa - F\sin(\theta + \alpha)\,\delta\theta\right] ds + c\,\theta_1\,\delta\theta_1 \tag{5}$$

where $\delta\theta$ is variation of $\theta$ and $\delta\kappa = -\dfrac{d\delta\theta}{ds}$. After integration by parts and using (4) for $\kappa$, we obtain

$$\delta\Pi = -EI\kappa\,\delta\theta\Big|_0^\ell - \int_0^\ell \left[EI\frac{d^2\theta}{ds^2} + F\sin(\theta + \alpha)\right]\delta\theta\, ds + c\,\theta_1\,\delta\theta_1 \tag{6}$$

By making $\delta\Pi = 0$, we obtain the differential equation

$$EI\frac{d^2\theta}{ds^2} + F\sin(\theta + \alpha) = 0 \tag{7}$$

24/12/2018 08:50

and boundary conditions. In our case these are

$$\kappa(0)=0, \quad -EI\kappa(\ell)+c\theta_1=0 \tag{8}$$

Thus, the cantilever equilibrium forms are solutions of second order ordinary differential equation (7) subject to boundary conditions (8).

The second variation of $\Pi$ as given by (3) is

$$\delta^2\Pi = -EI\delta\kappa\,\delta\theta\Big|_0^\ell - \int_0^\ell \left[EI\frac{d^2\delta\theta}{ds^2}+F\cos(\theta+\alpha)\delta\theta\right]\delta\theta\,ds + c(\delta\theta_1)^2 \tag{9}$$

The condition $\delta^2\Pi=0$ leads to the Jacobi accessory equation

$$EI\frac{d^2\delta\theta}{ds^2}+F\cos(\theta+\alpha)\delta\theta = 0 \tag{10}$$

and the following boundary conditions that are consistent with conditions (8)

$$\delta\kappa_0=0, \quad -EI\delta\kappa(s)+c\delta\theta(s)=0 \tag{11}$$

We recall that by the Jacobi test the equilibrium shape of the beam is unstable if any nontrivial solution of (10) under the boundary conditions (11) has a solution (conjugate points) in $0<s\leq\ell$.

**3 Solution**

In the following, we will use Jacobian elliptic functions $\text{sn}(x,k)$, $\text{cn}(x,k)$, $\text{dn}(x,k)$, Jacobi's epsilon function $\mathcal{E}(x,k)\equiv\int_0^x \text{dn}^2(t,k)dt$ and the complete elliptic integral of the first kind $K(k)$. Also, we will use the following derived Jacobian elliptic function $\text{sd}(x,k)\equiv\text{sn}(x,k)/\text{dn}(x,k)$ and $\text{cd}(x,k)\equiv\text{cn}(x,k)/\text{dn}(x,k)$ [36].

*3.1 Equilibrium*

We introduce the following non-dimensional parameters

$$\omega^2 \equiv \frac{F\ell^2}{EI}, \quad \beta^2 \equiv \frac{c\ell}{EI} \tag{12}$$

and from here on we use $\ell$ as a unit of length so we have $0 \leq s \leq 1$. We note that $\omega^2$ represent non-dimensional force. However, in diagrams and tables, we will use normalized force

$$\frac{F}{F_E} = \frac{\omega^2}{\pi^2} \qquad (13)$$

where $F_E \equiv \pi^2 \dfrac{EI}{\ell^2}$ is Euler critical force for buckling of a pin-ended column.

By using (12) the equation (7) and boundary conditions (8) become

$$\frac{d^2\theta}{ds^2} + \omega^2 \sin(\theta + \alpha) = 0, \qquad (14)$$

$$\kappa(0) = 0, \quad -\ell\kappa(1) + \beta^2 \theta_1 = 0. \qquad (15)$$

The solution of (14) is [14, 37-39]

$$\theta = -\alpha + 2\sin^{-1}\left[k\,\mathrm{sn}(\omega s + C, k)\right] \qquad (16)$$

where $C$ is a constant of integration and $k$ is the elliptic modulus. The base curve curvature is determined from (4)

$$\ell\kappa = -2k\Omega\,\mathrm{cn}(\omega s + C, k), \qquad (17)$$

Using (17) for $\kappa$ and (16) for $\theta$ we from the boundary conditions (15) obtain the relations

$$\mathrm{cn}(C, k) = 0 \qquad (18)$$

$$-\frac{\alpha}{2} + \sin^{-1}\left[k\,\mathrm{sn}(\omega + C, k)\right] + \frac{k\omega}{\beta^2}\mathrm{cn}(\omega + C, k) = 0 \qquad (19)$$

From these we find

$$C = K(k) \qquad (20)$$

$$-\frac{\alpha}{2} + \sin^{-1}\left[k\,\mathrm{cd}(\omega, k)\right] - \frac{k\sqrt{1-k^2}\,\omega}{\beta^2}\mathrm{sd}(\omega, k) = 0 \qquad (21)$$

In this way, we reduce the problem to solving the equation (21) for unknown $k$. This can be done numerically. Finally, substituting (16) for $\theta$ into (1) and perform integration, we obtain the coordinates of the points of the beam base curve

$$x = \xi \cos\alpha + \eta \sin\alpha, \qquad y = -\xi \sin\alpha + \eta \cos\alpha \tag{22}$$

where

$$\xi = \frac{2}{\omega}\Big[\mathcal{E}(\omega + C, k) - \mathcal{E}(\omega s + C, k)\Big] - (1 - s), \tag{23}$$

$$\eta = \frac{2k}{\omega}\Big[\operatorname{cn}(\omega s + C, k) - \operatorname{cn}(\omega + C, k)\Big]. \tag{24}$$

*3.2 Stability*

Using (12) we obtain from (10) and (11) the Jacobi's accessory equation in the following form ([35])

$$\frac{d^2\vartheta}{ds^2} + \omega^2 \cos(\theta + \alpha)\vartheta = 0 \tag{25}$$

where $\vartheta \equiv \delta\theta$. The corresponded boundary conditions (11) become

$$\frac{d\vartheta}{ds}(0) = 0, \qquad \left(\frac{d\vartheta}{ds} + \beta^2 \vartheta\right)(s_c) = 0. \tag{26}$$

The solution of (25) has the form [40]

$$\vartheta(s) = C_1 \vartheta_1(s) + C_2 \vartheta_2(s) \tag{27}$$

where $C_1$, $C_2$ are constant of integration and

$$\vartheta_1 \equiv \frac{\partial\theta}{\partial C} = 2k \operatorname{cn}(\omega s + C, k), \tag{28}$$

$$\vartheta_2 \equiv \frac{\partial\theta}{\partial k} = \frac{2}{1-k^2}\Big\{\operatorname{sn}(\omega s + C, k)\operatorname{dn}(\omega s + C, k) \\ -\Big[\mathcal{E}(\omega s + C, k) - (1-k^2)(\omega s + C)\Big]\operatorname{cn}(\omega s + C, k)\Big\}, \tag{29}$$

Substituting (27) into boundary conditions (26), we obtain a homogeneous system of equations for $C_1$ and $C_2$ which has a non-trivial solution if it's determinate vanish. This condition leads to the following equation for $s_c$

$$\left[\mathcal{E}\left(\omega s_c, k\right) - \left(1 - k^2\right)\omega s_c\right]\left[\operatorname{sn}\left(\omega s_c, k\right) + \frac{\omega}{\beta^2}\operatorname{cd}\left(\omega s_c, k\right)\right] + \operatorname{cn}\left(\omega s_c, k\right)\operatorname{dn}\left(\omega s_c, k\right) - \left(1 - k^2\right)\frac{\omega}{\beta^2}\operatorname{sn}\left(\omega s_c, k\right) = 0 \quad , \quad (30)$$

where we omit the factor $k\omega$. By Jacobi's test [35], the necessary condition for $\delta^2\Pi > 0$ is that the smallest root of this equation is $s_c > 1$. Therefore, on the other hand, if $0 < s_c \leq 1$ then the beam shape is unstable. We note that $\alpha$ dropped from the stability analysis. This should be clear from the expressions for the beam coordinates (22): $\alpha$ affect only the rod position but not its shape.

As a verification of the above equations, we consider the case $\beta^2 = \infty$ (clamped cantilever). In this case, the equation (21) reduces to well-known $\operatorname{cn}\left(\omega, k\right) = 0$ , so $\omega = \left(2n - 1\right)K\left(k\right)$ where $n$ is an integer. Also, the equation (30) becomes the equation for cantilever given in [40] (Eq 21 therein. In this equation first sign – should be +).

## 4 Examples

With the above solution, we can easily construct various bifurcation diagrams, load-deflection diagrams and calculate a deformed beam shape. The stability of the beam shapes can be treated by a numerical solution of (30) using the procedure described in [41]. For all numerical calculations with elliptic functions we use Elfun18 library [42].

To verify the present solution, we compare our calculations of the beam free end coordinates, and the tangent angle at the beam ends with those from [18]. The results are given in Table 1 where we can observe scatter but acceptable difference within 10% in all cases except that of $\omega^2 = 1$ where the difference is up to 20%. The comparison is also shown in Fig 2.

**Table 1**. Numerical values for column shapes shown in Fig 3. $\beta^2 = 34.69$ , $\alpha = \pi/4$

| $\dfrac{F\ell^2}{EI}$ | [18] | | | | Present | | | | Relative difference % | | | |
|---|---|---|---|---|---|---|---|---|---|---|---|---|
| | $x_0/\ell$ | $y_0/\ell$ | $\theta_0^0$ | $\theta_1^0$ | $x_0/\ell$ | $y_0/\ell$ | $\theta_0^0$ | $\theta_1^0$ | $x_0/\ell$ | $y_0/\ell$ | $\theta_0^0$ | $\theta_1^0$ |
| 1 | 0.951 | 0.258 | 25.1 | 1.2 | 0.93611 | 0.32338 | 28.013 | 1.471 | 1.6 | -20.2 | -10.4 | -18.4 |
| 2 | 0.750 | 0.601 | 59.5 | 3.0 | 0.72701 | 0.62366 | 59.184 | 3.155 | 3.2 | -3.6 | 0.5 | -4.9 |
| 2.9 | 0.561 | 0.752 | 84.0 | 4.8 | 0.53216 | 0.75927 | 78.919 | 4.374 | 5.4 | -1.0 | 6.4 | 9.7 |
| 6.8 | 0.099 | 0.856 | 114.8 | 7.5 | 0.09051 | 0.86115 | 113.691 | 7.558 | 9.4 | -0.6 | 1.0 | -0.8 |
| 11.2 | -0.105 | 0.842 | 126.6 | 10.2 | -0.10100 | 0.84792 | 125.142 | 9.770 | 4.0 | -0.7 | 1.2 | 4.4 |

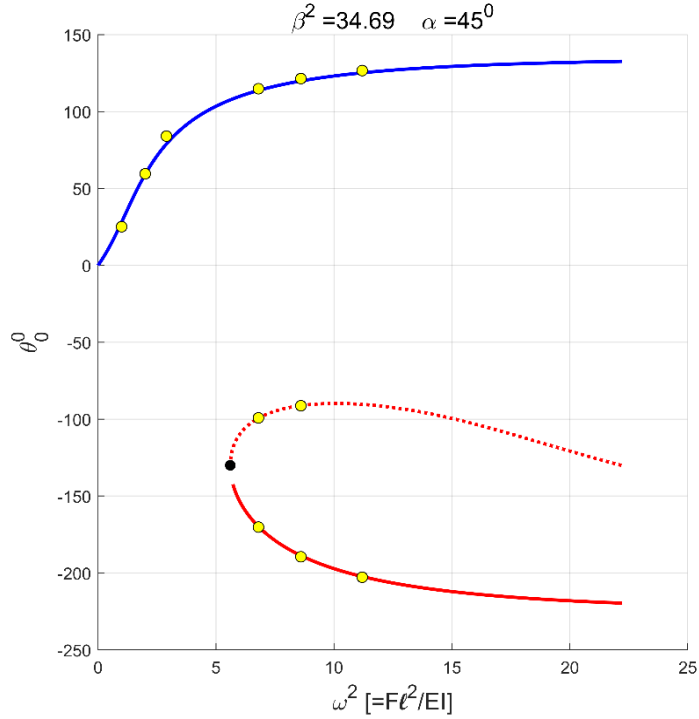

**Figure 2**. Free end tangent angle as a function of normalized force. Dotted line represent an unstable solution branch. Bright dots are values from [18]. Critical normalized force is 5.6071, corresponding free end angle is -130.012° (black dot)

As an example of the application of the present solution, we consider the case $\beta^2 = \pi^2/2$ and $\alpha = \pi/4$. The bifurcation diagram for the case is shown in Fig 3. Each branch start where $\frac{dk}{d\omega} = 0$, or, using (21),

$$\left(1 + \frac{1}{\beta^2}\right)\operatorname{sn}(\omega, k) + \frac{\omega}{\beta^2}\operatorname{cd}(\omega, k) = 0, \tag{31}$$

The start point of branch is thus the solution of the system of equations (21) and (31). For each branch we have two solutions, i.e., each branch is split into two parts one for $k > 0$ (upper) and one for $k < 0$ (lower). We see from the figure that only the first branch can be stable; all other branches are unstable. The upper part of the first branch that emerges from the initial beam straight state is entirely stable (Fig 4). The other, lower part, can be reached only by applying a force higher than a critical one to some pre-deformed shape (Fig 5). Note, that this part is unstable from point A to B (see Fig 3). The stationary point of the lower part is where $\frac{d\omega}{dk} = 0$. This condition, using (21), leads to an equation which the same as equation (30) for $s_c = 1$, i.e., the stationary point lie on the boundary of the stable region. We thus obtain the critical value of $\omega$ and $k$ by a solution of the system of equations (21) and (30).

We obtain a similar behavior also for other values of $\beta^2$ and $\alpha$, and for $\beta^2 = \infty$ we obtain the solution for the clamped beam which is discussed in [41]. Thus, we conclude that in general, only the first branch of the spring-hinged cantilever beam can be stable, all others are unstable.

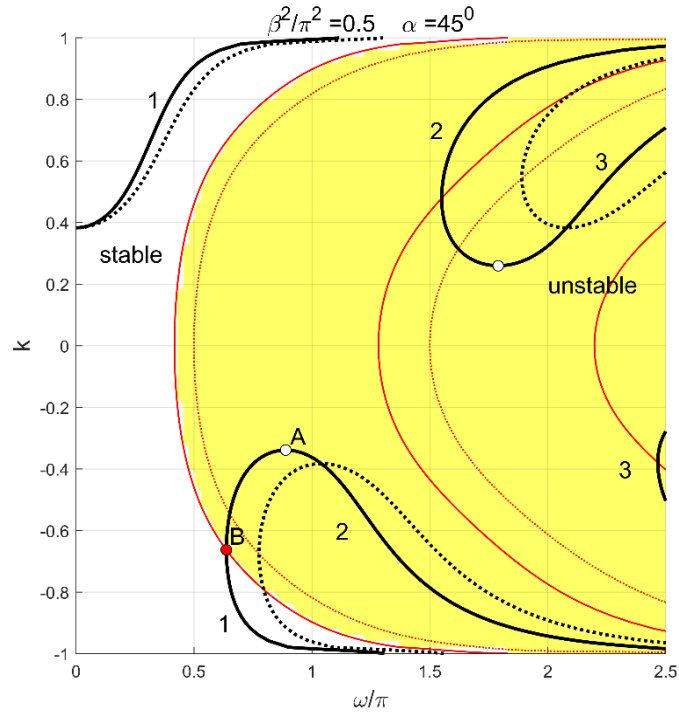

**Figure 3.** Bifurcation diagram with two solution branches. All dotted lines are for the clamped cantilever beam. Tiny lines are for $d\omega/dk = 0$. The start point A of the lower part of the first branch is at $(0.88888, -0.33870)$. This part of the branch becomes stable after passing the point B which is at $(0.63713, -0.66276)$.

**Table 2.** Numerical values for the beam shapes shown in Fig 4. $\beta^2 = \pi^2/2$, $\alpha = \pi/4$

| $\omega/\pi$ | $k$ | $F/F_E$ | $x_0/\ell$ | $y_0/\ell$ | $\theta_0^0$ | $\theta_1^0$ |
|---|---|---|---|---|---|---|
| 0.25 | 0.55357 | 0.0625 | 0.95367 | 0.28915 | 22.224 | 6.294 |
| 0.5 | 0.90653 | 0.25 | 0.38751 | 0.87127 | 85.062 | 25.499 |
| 0.75 | 0.98578 | 0.5625 | -0.06172 | 0.92742 | 115.655 | 39.457 |
| 1 | 0.99770 | 1 | -0.27987 | 0.88885 | 127.221 | 49.345 |

**Table 3**. Numerical values for column shapes shown in Fig 5. $\beta^2 = \pi^2/2$ , $\alpha = \pi/4$

| $\omega/\pi$ | $k$ | $F/F_E$ | $x_0/\ell$ | $y_0/\ell$ | $\theta_0^0$ | $\theta_1^0$ |
|---|---|---|---|---|---|---|
| 0.63714 | -0.66642 | 0.40595* | -0.06048 | -0.86397 | -128.583 | -30.412 |
| 0.75 | -0.92533 | 0.5625 | -0.60967 | -0.49959 | -180.436 | -50.558 |
| 1 | -0.99022 | 1 | -0.77905 | -0.09016 | -208.963 | -70.431 |

* the critical force.

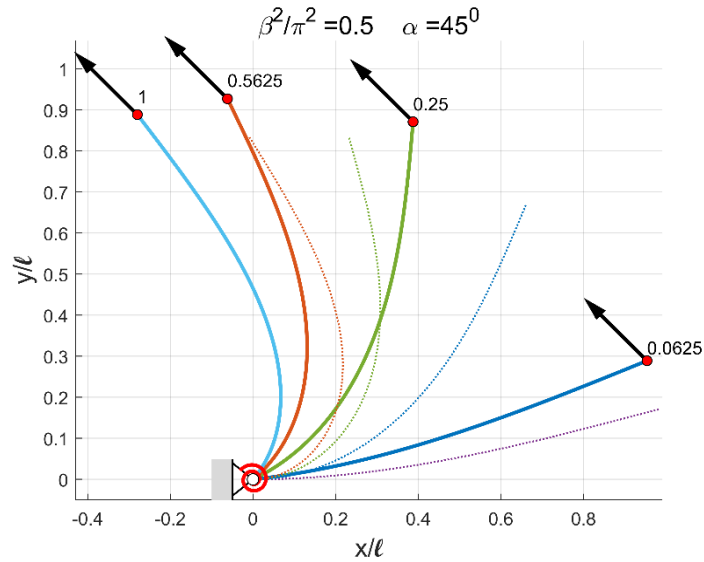

**Figure 4.** Stable equilibrium shapes for various values of $F/F_E$. Dotted lines correspond shapes for the clamped cantilever.

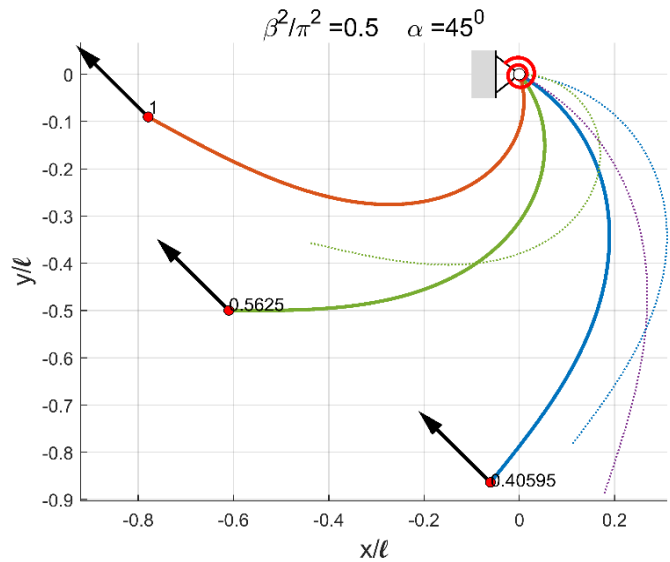

**Figure 5.** Stable equilibrium shapes for various values of $F/F_E$. Dotted lines correspond forms for the clamped cantilever.

24/12/2018 08:50

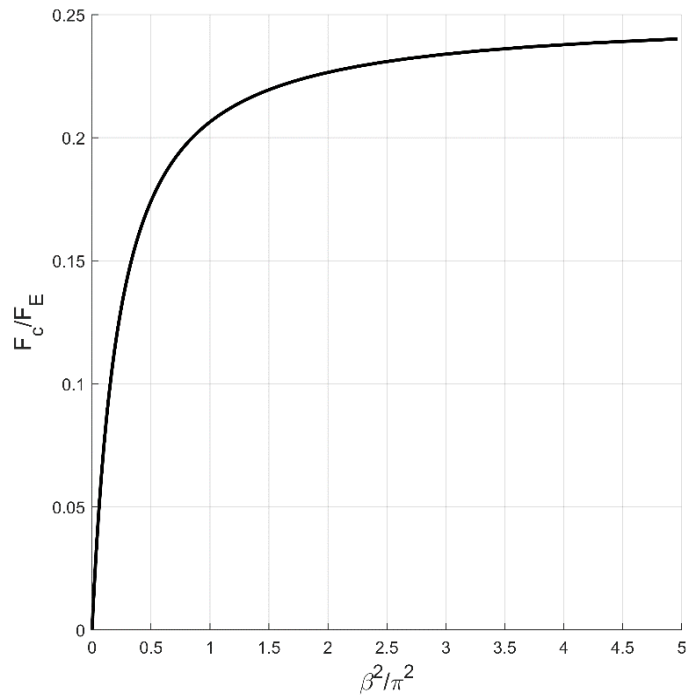

**Figure 6.** Critical force $F_c/F_E$ versus spring stiffness $\beta^2$. $F/F_E \to 1/4$ as $\beta^2 \to \infty$.

For the last example, we consider the spring-hinged cantilever beam under pure compression. From the graph of the critical force in Fig 6, we see that the effect of spring become relatively small for say $\beta^2 > 40$ wherever it becomes less than 5% of critical force for the clamped beam. We can see from the graph in Fig 7 that after buckling, the beam continues to support load, i.e., the force still increases with increasing deflection. From the bifurcation diagram in Fig 8, we see that only the first buckled form is stable. All other shapes are unstable. Some stable shapes are shown in Fig 9.

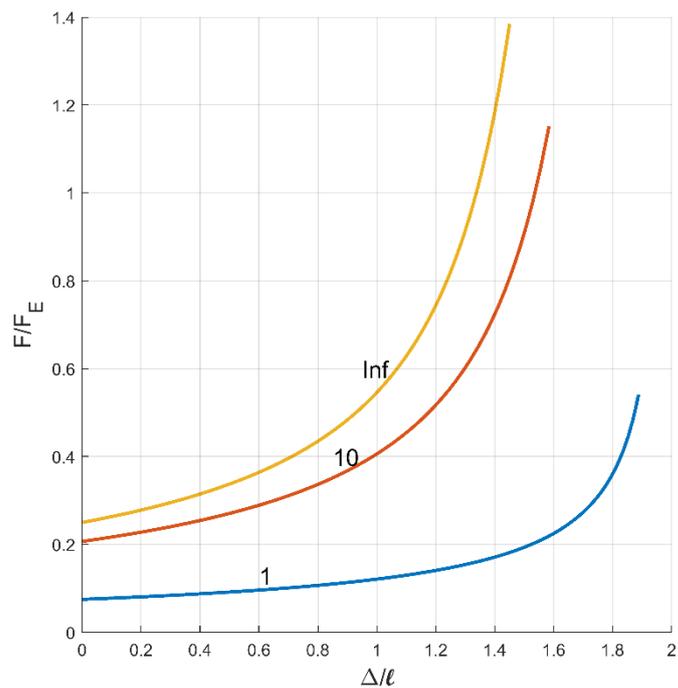

**Figure 7.** Load-deflection diagram for beam under compression for various values of normalized spring stiffness $\beta^2$.

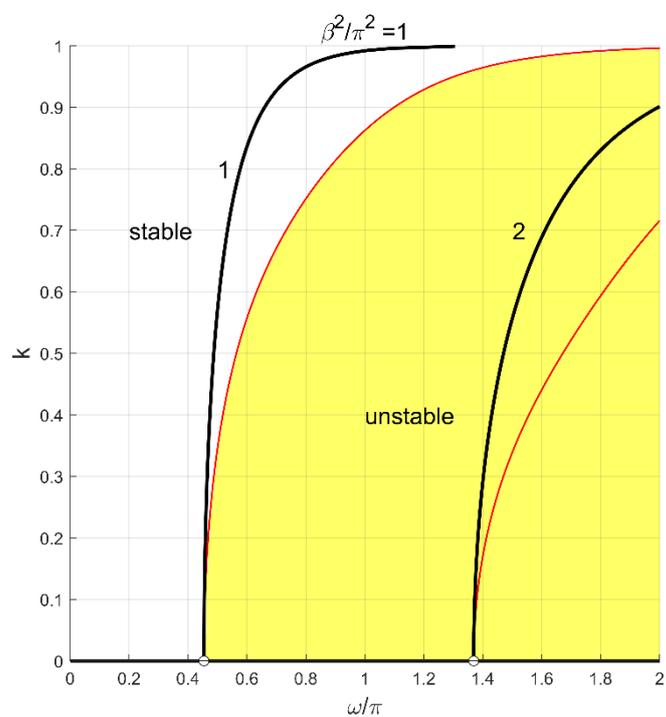

**Figure 8.** Bifurcation diagram with the first two branches for the cantilever beam under pure compression.

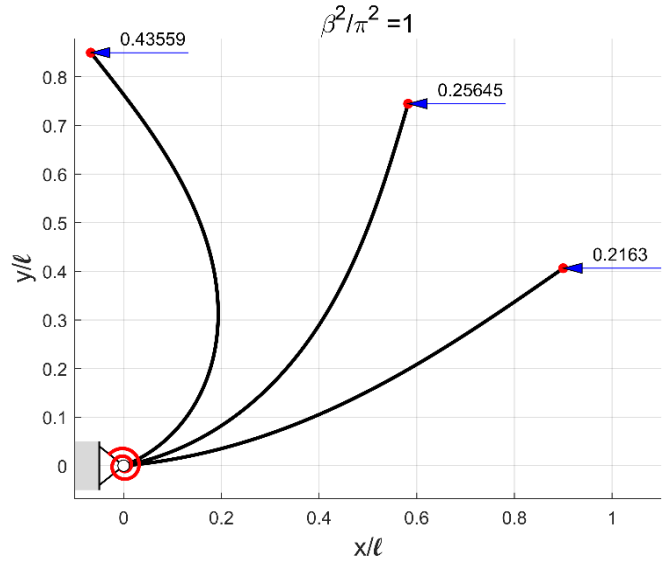

**Figure 9.** Equilibrium shapes for various values of $F/F_E$ for data given in Table 4.

**Table 4.** Numerical values for column shapes shown in Fig 9. $\beta^2 = \pi^2$.

| $k$ | $\omega$ | $F/F_E$ | $\Delta/\ell = 1 - x_0/\ell$ | $y_0/\ell$ |
|---|---|---|---|---|
| 0 | 1.427187* | 0.206377 | 0 | 0 |
| 0.3 | 1.461077 | 0.216295 | 0.100091 | 0.406231 |
| 0.6 | 1.590929 | 0.256449 | 0.417597 | 0.744692 |
| 0.9 | 2.073437 | 0.435594 | 1.067300 | 0.849782 |

\* critical value

Trough we demonstrate the stability of the beam with the graphs, two cases can be treated analytically. The first case is a straight beam and the second is the case of small deflections.

The solution of (21) corresponds to the straight form is $k = 0$. In this case we from (16), (17), (23), (24) obtain.

$$\theta = 0, \quad \kappa = 0, \quad x = 1 - s, \quad y = 0 \tag{32}$$

To determine straight form stability, we consider (30) which for $k = 0$ reduce to

$$-\frac{\omega}{\beta^2}\sin(\omega s_c) + \cos(\omega s_c) = 0 \tag{33}$$

The solution to this equation is $s_c = \frac{1}{\omega} \tan^{-1}\left[\frac{\beta^2}{\omega}\right]$. For $s_c = 1$ it becomes the well-known characteristic equation for the critical value $\omega_c$ of $\omega$ [16]

$$\beta^2 = \omega_c \tan \omega_c \tag{34}$$

Thus, the straight beam is stable for $\omega < \omega_c$. In particular case $\beta^2 = 0$ (no spring) then $\omega_c = 0$, i.e., the straight beam is unstable. If $\beta^2 = \infty$ (clamped end) then $\omega_c = \pi/2$. Graph of (34) is shown in Fig 7.

For small $k$, i.e., small deflection, we obtain from (20), (21)

$$C = \frac{\pi}{2} + O\left(k^2\right) \tag{35}$$

$$\frac{\omega}{\omega_0} = 1 + \frac{\omega_2}{12} k^2 + O\left(k^4\right) \tag{36}$$

where

$$\omega_2 = \frac{3\omega_0^4 + \left(5 + 6\beta^2\right)\beta^2 \omega_0^2 + 3\beta^6 \left(1 + \beta^2\right)}{\left(\omega_0^2 + \beta^4\right)\left(\omega_0^2 + \beta^4 + \beta^2\right)} \tag{37}$$

and $\omega_0$ is solution of

$$-\frac{\omega_0}{\beta^2} \sin \omega_0 + \cos \omega_0 = 0 \tag{38}$$

Future from (16), (17), (23), (24) we have

$$\theta = 2k \cos\left(\omega_0 s\right) + O\left(k^3\right) \tag{39}$$

$$\kappa = 2k\omega_0 \sin\left(\omega_0 s\right) + O\left(k^3\right) \tag{40}$$

$$x = 1 - s + O\left(k^2\right) \tag{41}$$

$$y = \frac{2k}{\omega_0}\left[\sin \omega_0 - \sin\left(\omega_0 s\right)\right] + O\left(k^3\right) \tag{42}$$

To assess the stability of small deflection, we substitute $s_c = 1$ into (30) and expand it into a power series of $k$. In this way we obtain

24/12/2018 08:50

$$\frac{\omega_c}{\omega_0} = 1 + \frac{\omega_2}{4} k^2 + O(k^4) \tag{43}$$

Comparing (36) and (43) we find that $\omega < \omega_c$, i.e., the initial deflected form is stable.

## 5. Follower force

We obtain a solution for large deflection of the rod under follower force by setting [21]

$$\alpha = \gamma - \theta_0 \tag{44}$$

where $\theta_0$ is the free end tangent angle, and $\gamma$ is the angle between force and base curve tangent at the rod free end. Substituting this into the expression (16) for $\theta$ and set $s = 0$ yield

$$k = \sin\frac{\gamma}{2} \tag{45}$$

Substituting (44) into the characteristic equation, (21) we obtain the following formula for free end tangent angle

$$\theta_0 = \gamma - 2\sin^{-1}\left[k\,\mathrm{cd}(\omega,k)\right] + \frac{2k\sqrt{1-k^2}\,\omega}{\beta^2}\,\mathrm{sd}(\omega,k) \tag{46}$$

For each $\omega$ and $\gamma$ we can thus calculate $k$ from (45), $\theta_0$ from (46) and $\alpha$ from (44). In words: the problem has a unique solution, i.e., an equilibrium form of the spring-hinged cantilever under a follow force is unique. This generalizes results given in [21] where $\beta^2 = \infty$ and $\gamma = 0$.

The angle $\theta_0$ reaches stationary value when $\dfrac{d\theta_0}{d\omega} = 0$. From this condition, we deduce the following equation

$$\mathrm{sd}(\omega,k) + \omega\,\mathrm{cd}(\omega,k) + \beta^2\mathrm{sn}(\omega,k) = 0. \tag{47}$$

For example, when $\beta^2 = \infty$ then (47) reduces to $\mathrm{sn}(\omega,k) = 0$. The smallest positive root of this equation is $\omega = 2K(k)$. For $\gamma = \pi/2$ this gives $\omega = 3.70815$ or $F_c = 13.75037$. This value differs from Shvartsman's [25] by less than 0.3%. We note that when $\omega = 2K(k)$ then $\mathrm{cn}(\omega,k) = -1$ and $\mathrm{dn}(\omega,k) = 1$ so (46) reduce to $\theta_0 = 2\gamma$, i.e., value which is independent of $\beta^2$. This can be observed in Fig 10. To obtain the value of $\beta^2$ for which $\theta_{0,\max} = 2\pi$ we substitute $\beta^2$ from (47) into (46). A solution of the resulting equation gives $\omega = 2.52909$ and thus $\beta^2 = 0.74324$.

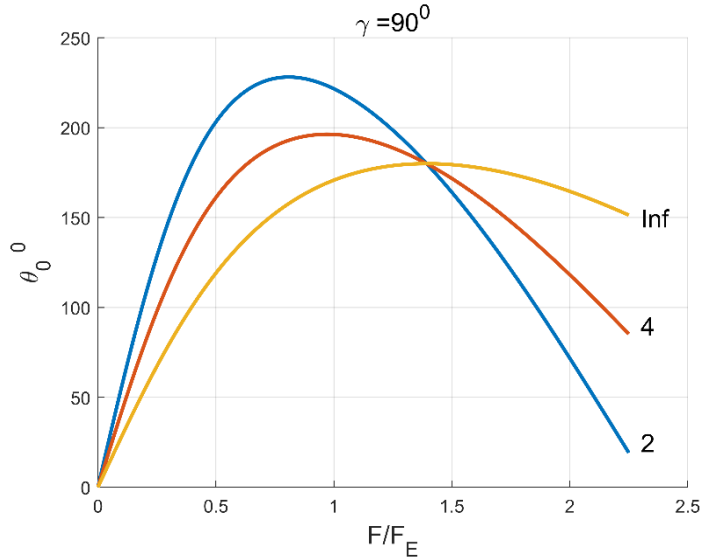

**Figure 10.** Free end tangent angle as a function of normalized follower force for various values of normalized spring stiffness $\beta^2$. $F/F_E \to 0.64807$ as $\beta^2 \to 0$

**Table 5.** Numerical values for beam shapes shown in Fig 11. $\beta^2 = \pi^2/2$  $\gamma = \pi/2$

| $\omega/\pi$ | $k$ | $F/F_E$ | $x_0/\ell$ | $y_0/\ell$ | $\theta_0^{\,0}$ | $\theta_1^{\,0}$ |
|---|---|---|---|---|---|---|
| 0.25 |  | 0.0625 | 0.94249 | 0.32144 | 24.710 | 7.094 |
| 0.5 | $\sqrt{2}/2$ | 0.25 | 0.33137 | 0.87781 | 92.127 | 24.778 |
| 0.75 |  | 0.5625 | -0.34491 | 0.68558 | 163.135 | 34.097 |
| 1 |  | 1 | -0.28124 | 0.48360 | 191.425 | 20.612 |

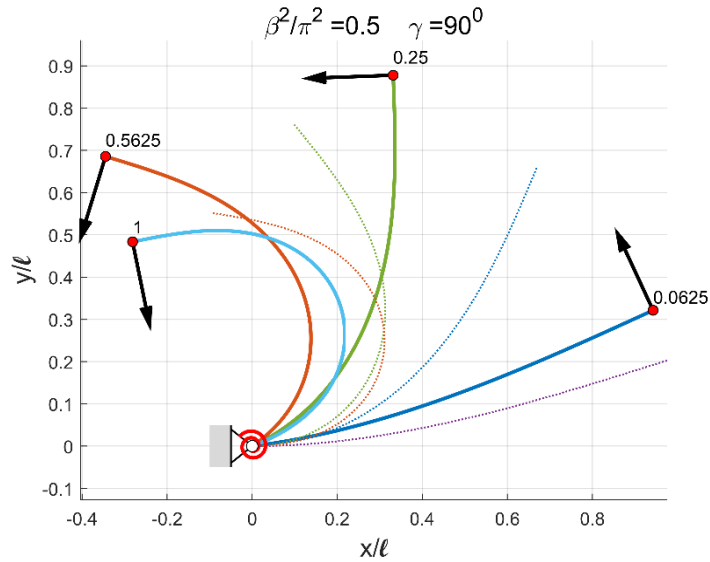

**Figure 11**. Equilibrium shapes of a spring-supported cantilever for various values of follower force. Dotted lines are for the corresponded clamped cantilever beam.

**Conclusions**

For the spring-hinged cantilever beam under a pure compression force, we prove that its first buckled mode is unconditionally stable in the sense that the beam after buckling retains its loading capacity; all higher buckled modes are unstable. The spring-hinged cantilever beam under inclined force has two stable equilibrium solutions. One completely stable emerges from the initial state, and the other, partly stable is reached from a pre-deformed state. We also give an analytical solution for the cantilever subject to a follower force.

In the end, we add that an advantage of a closed form analytical solution of the problem, compared to other methods, is that we have on our disposal an exact 'big picture' of the solution, i.e., whole phase plane on which we can relatively easily determine equilibrium conditions and its stability.

24/12/2018 08:50

24/12/2018 08:50

24/12/2018 08:50